\documentclass[aps,prb,twocolumn,groupedaddress]{revtex4}

\usepackage{epsfig}

\begin{document}


\title{Transport through the Interface between a Semiconducting Carbon Nanotube and a Metal Electrode}

\author{Takeshi Nakanishi, Adrian Bachtold, and Cees Dekker}
\affiliation{Department of Applied Physics and DIMES, Delft University of Technology,
Lorentzweg 1, 2628 CJ Delft, The Netherlands}


\date{\today}

\begin{abstract}
We report a numerical study of the tunnel conductance through the Schottky barrier at the contact between a semiconducting carbon nanotube and a metal electrode. 
In a planar gate model the asymmetry between the p--doped and the n--doped region is shown to depend mainly on the difference between the electrode Fermi level and the band gap of carbon nanotubes.
We quantitatively show how the gate/nanotube distance is important to get large on--off ratios.
We explain the bend of the current versus gate voltage as the transition from a thermal--activation region to a tunneling region.
A good agreement is obtained with experimental results for carbon nanotubes field--effect transistors.
\end{abstract}

\pacs{}

\maketitle

Recently, we have reported logic circuits with carbon nanotube transistors.\cite{Bachtold_et_al_2001a}
Single--wall carbon nanotubes (CNs) are novel quantum wires consisting of tubular graphite sheets,\cite{Iijima_1991a} which can be synthesized in structures $\sim1$ nm in diameter and microns long.\cite{Iijima_et_al_1993a, Bethune_et_al_1993a}
The one--dimensional (1d) nature of CNs requires careful analysis with respect to the screening and the band bending near the contact to the electrode.
The purpose of this study is to calculate the transmission probability through the Schottky barrier at the interface between a semiconducting CN and a metal electrode and to compare it with the experiments.

The contact between metal electrodes and CNs has been studied theoretically before.\cite{Kong_et_al_1999a,Choi_et_al_1999a,Rochefort_et_al_2001a,Anantram_et_al_2000a,Nakanishi_and_Ando_2000a,Odintsov_2000a,Odintsov_and_Tokura_2000a,Leonard_and_Tersoff_1999a,Leonard_and_Tersoff_2000a,Tersoff_1999a,Xue_and_Datta_1999a}
The transmission at a contact has been evaluated by {\it ab initio} calculation\cite{Kong_et_al_1999a,Choi_et_al_1999a,Rochefort_et_al_2001a} as well as by using simple models.\cite{Anantram_et_al_2000a,Nakanishi_and_Ando_2000a,Odintsov_2000a}
Using the latter, it has been shown that the charge transfer in CNs is dramatically different than in 2d or 3d systems and that this strongly modifies the band bending near the contact.\cite{Odintsov_2000a, Odintsov_and_Tokura_2000a}
The band bending varies slowly with a logarithmic dependence on the distance in a coaxial gate model where the gate electrode is a cylinder surrounding the CN.\cite{Odintsov_2000a,Odintsov_and_Tokura_2000a}
The variation in distance depends on the gate voltage, but is on the order of the radius of the cylinder.
For the same reason, p--n junctions of 1d CNs have been shown to behave differently than conventional p--n junctions.\cite{Leonard_and_Tersoff_1999a}

Experiments on samples with semiconducting tubes acting as field--effect transistors have been reported previously.\cite{Tans_et_al_1998a, Martel_et_al_1998a}
The Schottky barrier was studied by transport\cite{Zhou_et_al_2000a} and scanning probe experiments\cite{Freitag_et_al_2001a}.
Typically, an oxidized Si wafer was used as the gate.
The distance between the gate and the CN typically was long, on the order of the CN length.
Transconductance measurements showed that the tubes were p--doped, and that the capacitive coupling between the CN and the gate was too weak to n--dope the CN.
Recently, various groups have improved the sample device and have reported strong doping from n-- to p--type by using a very close gate of oxidized Al,\cite{Bachtold_et_al_2001a} large diameter nanotubes, \cite{Park_and_Mceuen_2001a,Javey_et_al_2002a} or annealed nanotubes.\cite{Derycke_et_al_2001a,Martel_et_al_2001a}

In the present work, we report calculation of the conductance as a function of the gate voltage in a semiconducting nanotube.
We compare this with experimental data.
The sample layout is characterized by a planar gate that lies very close to the CN.
The gate thus strongly screens the Coulomb interaction between the electrons in the CN.
We compare the measurements with a semi-classical model based on Poisson's equation.
Our method of calculation is similar to that in ref. \cite{ Odintsov_and_Tokura_2000a}, but our model incorporates a planar gate and a CN connecting strongly to a bulk electrode, instead of the coaxial gate, \cite{Odintsov_2000a,Odintsov_and_Tokura_2000a} 1d electrode\cite{Rochefort_et_al_2001a}, or weakly contacted CN at the interface.\cite{Kong_et_al_1999a,Choi_et_al_1999a,Anantram_et_al_2000a,Nakanishi_and_Ando_2000a}
By using this realistic model, we evaluate the current versus gate voltage and compare that quantitatively to the real device.
A good match is obtained.
The numerical results of our calculation are shown in detail.
Finally, we discuss how we can improve the device in order to get better transistors characteristics by changing the material and the geometry of the device.

In our devices, the gate consists of a microfabricated Al wire with a well-insulating native Al$_2$O$_3$ layer, which lies beneath a semiconducting CN that is electrically contacted to two Au electrodes on CN.
In this configuration, the Al$_2$O$_3$ layer of a few nm thickness is much shorter than the separation between the contact electrodes ($\sim$ 100 nm).
Capacitance measurements on two large Al films separated by the same aluminum oxide layer gives a thickness of 2 nm, whereas ellipsometry measurements give a value of around 5 nm.
\begin{figure}[t]
\centerline{\epsfig{figure=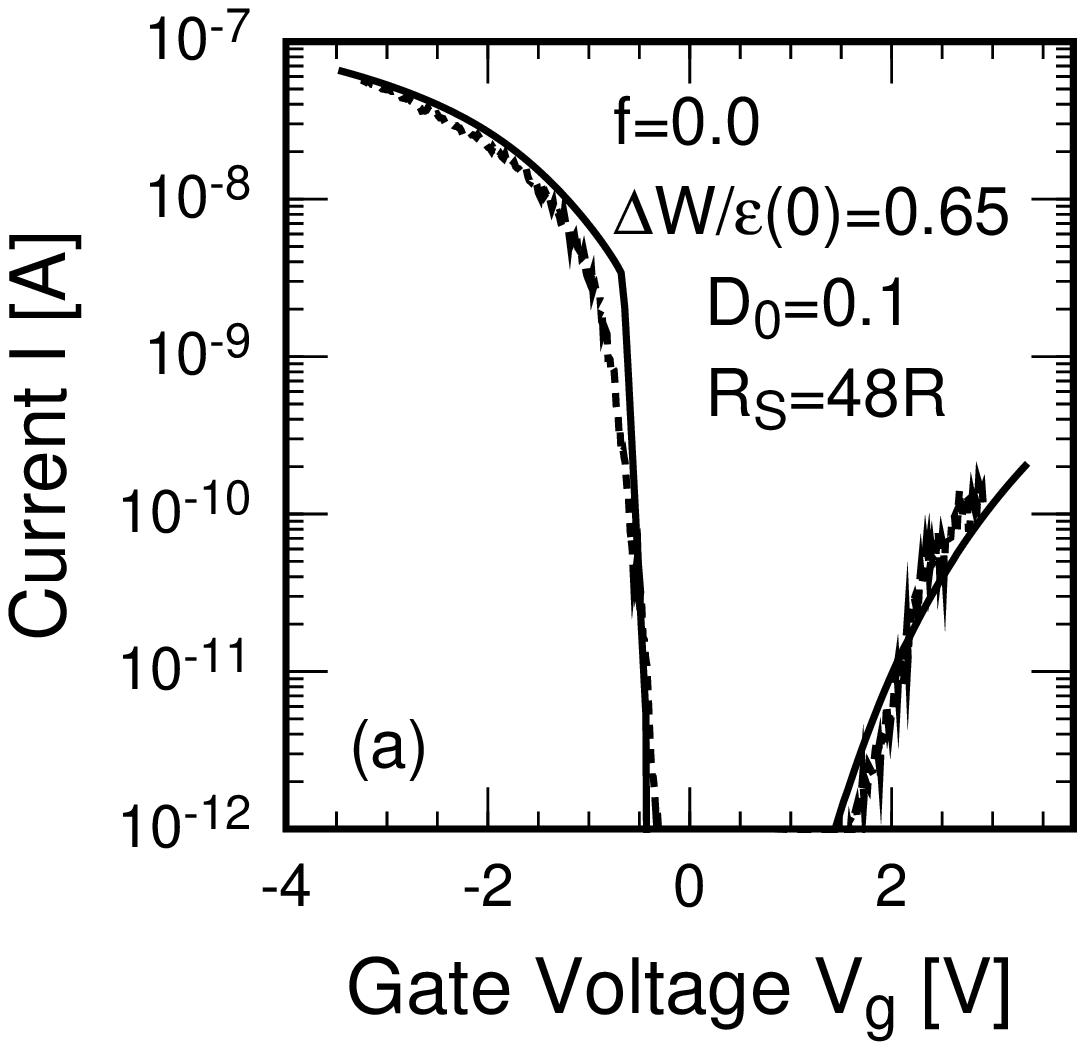,width=4.0cm}\epsfig{figure=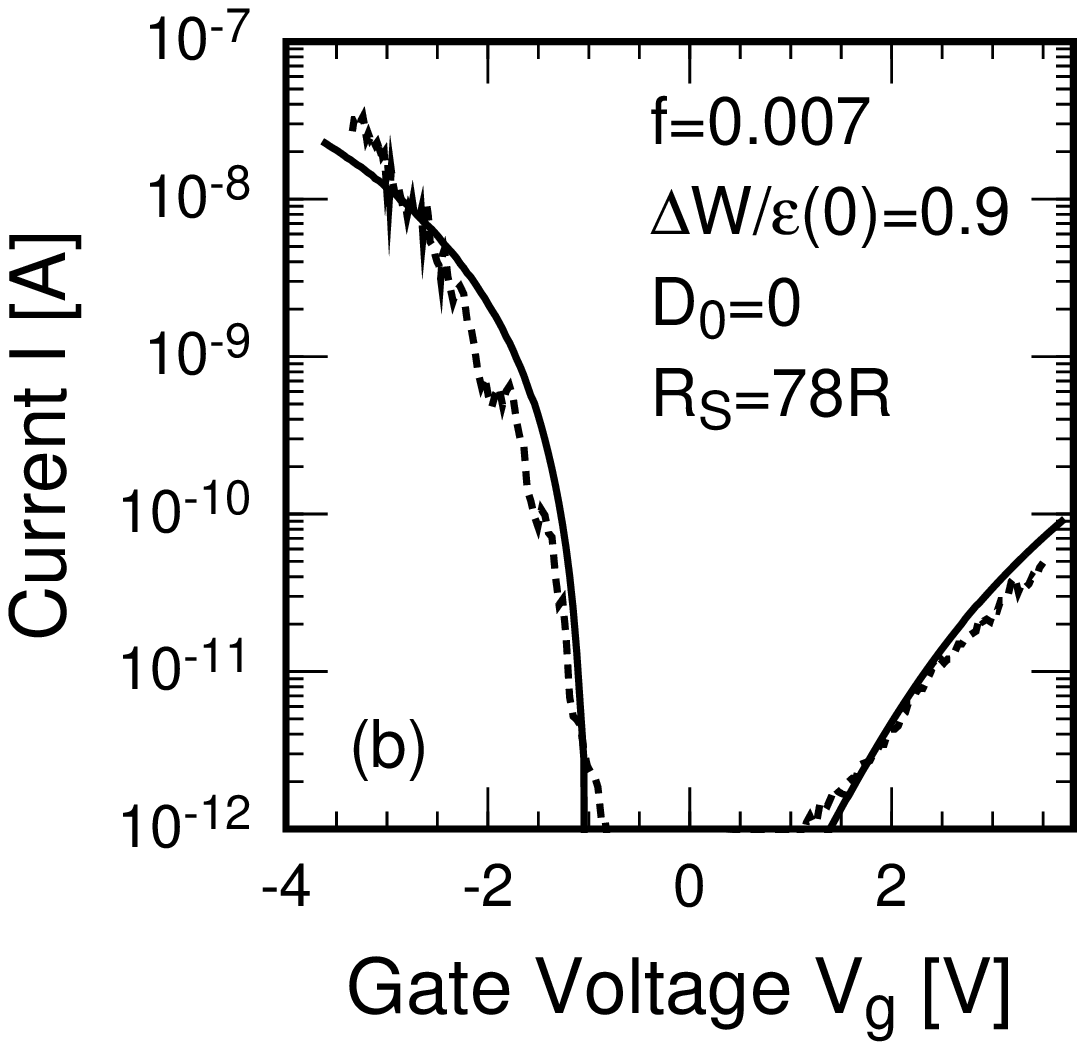,width=4.0cm}}
\centerline{\epsfig{figure=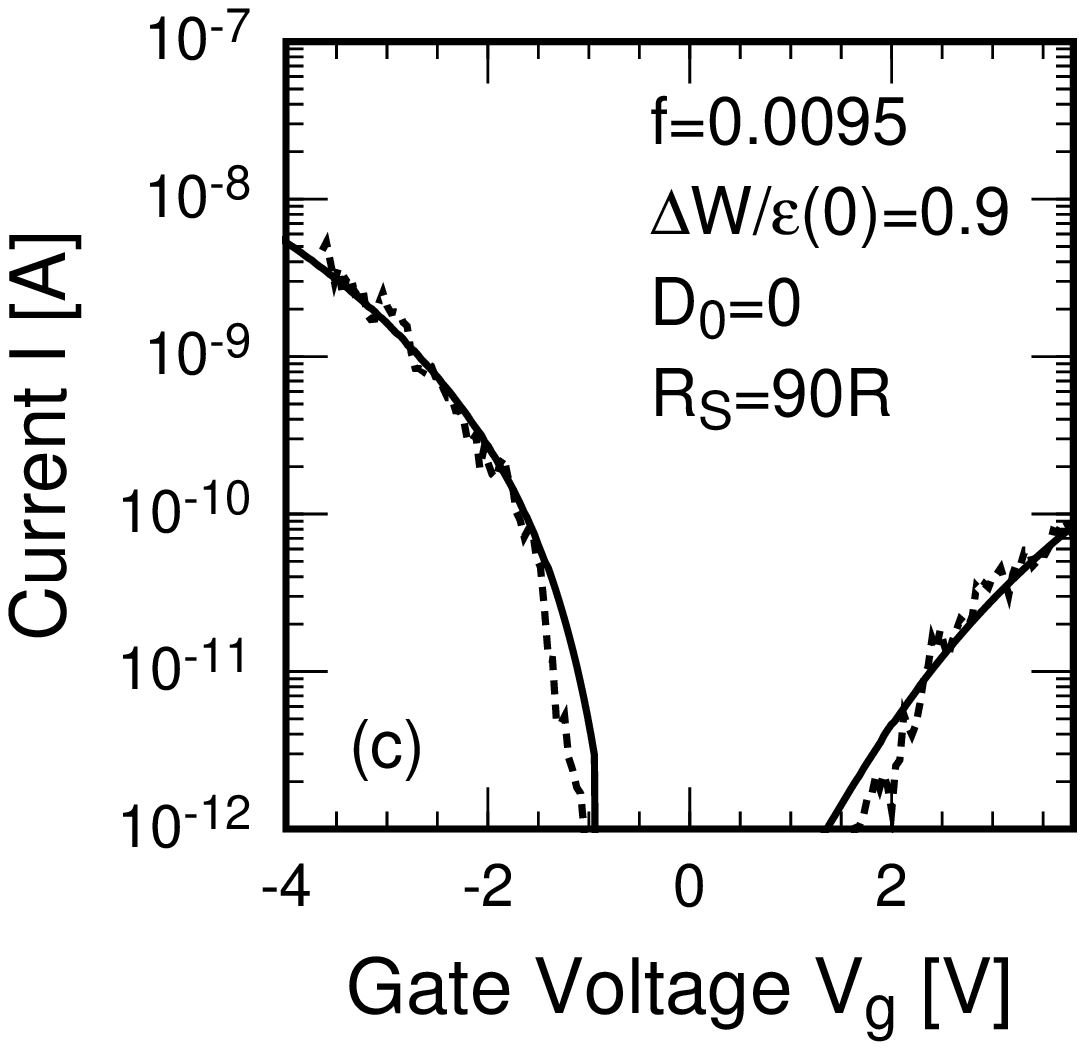,width=4.0cm}\raisebox{3cm}{\hspace{3cm}(d)}\hspace{-33mm}\raisebox{1cm}{\epsfig{figure=plane1.eps,width=4.0cm}}}
\caption{(a)-(c) Experimental $I(V_g)$ data for three typical nanotube transistors.
The experimental results are shown by dashed lines.
The results from the calculation are shown by solid lines.
(d) Model of the semi--infinite planar gate at $z=0$ and $y>0$.
The source ( or drain ) electrode is an infinite plane at $y=0$.
\label{Model}}
\end{figure}
Figures \ref{Model} (a), (b) and (c) show measurements of the current as a function of the gate voltage for three different samples.
The bias voltage is $V=5$ mV and the measurements are done in vacuum and at room temperature.
Strong doping from p to n is achieved for all samples.
Interestingly, an asymmetry is observed.
There are some sample dependences: We can see a smaller on--off ratio of $10^{4}$ in Figs.\ \ref{Model} (b) and (c), compared to the large on--off ratio $10^{5}$ in Fig.\ \ref{Model} (a).
The asymmetry of Fig. \ref{Model} (a) is stronger than Fig. \ref{Model} (c).
The current in the valence band is gradually decreasing and then rapidly decreasing in Fig.\ \ref{Model} (a), but the decrease is more slowly in Figs.\ \ref{Model} (b) and (c).
We will fit these experimental data by numerical calculations based on our model and explain how these sample dependence can be understood from our model.

We assume an end--bonded contact model in the calculation, which means that the electrons tunnel in the end of the tube from the metal contact.
Indeed, Bockrath {\it et al.} have shown that the CN is cut into segments when the electrode is patterned on top of the CN and that transport involves tunneling into the ends of the nanotube.\cite{Bockrath_et_al_1999a}
We used a planar--gate model as shown in Fig.\ \ref{Model} (d).
A CN is surrounded by dielectric material of dielectric constant $\kappa=5$, with a distance $R_s$ from the gate.
The Poisson equation relates the potential $\phi(y)$ at the surface of CN to the 1d charge density $\rho(y)$ in the CN and the potential $\Psi(y)$ of the gate,
\begin{eqnarray}
\phi_q&=&\phi_q^{(\rho)}+\phi_q^{(g)},\label{potential}\\
\phi_q^{(\rho)}&=&U_q\rho_q,\label{phi_rho}\\
\phi_q^{(g)}&=&M_q\Psi_q,\label{phi_g}
\end{eqnarray}
with the Fourier transformation $\phi_q$ and $\Psi_q$ of the potentials $\phi(y)$ and $\Psi(y)$, respectively, and,
\begin{eqnarray}
U_q&=&\frac{2}{\kappa}\left\{I_0(qR)K_0(qR)
-K_0(2qR_s)\right\},\label{eqn:chargekernel}
\\
M_q&=&\exp[-|q|R_s],
\label{eqn:gatem}
\end{eqnarray}
where $I_0$ and $K_0$ are the modified Bessel functions.
The first term of the kernel Eq. (\ref{eqn:chargekernel}) describes the self--capacitance within the tube.
The second term is the term of mutual capacitance, {\it i.e.}, the coulomb potential due to the charge on the gate, which is induced by the charge on the CN.
The analytic expression is a good approximation for the case $R_s\gg R$ with a radius $R$.
In calculation of the mutual term, we approximately presume the CN to be a wire, instead of taking into account the dependence on $z$ of the charge density and the potential $\phi$ on the surface of CN.
Furthermore, the interface between the insulator and vacuum in the experiment is ignored in the model.
We can neglect the correction of an image charge in the insulator at the opposite side of CN in the second term of Eq. (\ref{eqn:chargekernel}), because it is shown in the calculation that the second term is much smaller than the first term.

In equilibrium, the charge density is related to the energy $\bar{E}_0(y)=E_0(y)-E_{\rm F}$ of the charge neutrality level $E_0$ of the CN measured from the Fermi level $E_{\rm F}$.
We obtain
\begin{equation}
\rho(y)=\rho_0(y)+\frac{8\pi R}{\sqrt{3} a^2}e\left(D_0 \bar{E}_0(y)e^{-y/d}+\frac{f}{\kappa}\right)\mbox{sign}(y),
\label{Total Charge Density}
\end{equation}
with the lattice constant $a$, the pinning strength $D_0$, an effective decay constant $d$ of the surface states that equals about $2$ nm,\cite{Leonard_and_Tersoff_2000a} the doping fraction $f$ (the number of doped charge carriers per atom of the CN)\cite{Leonard_and_Tersoff_1999a}, and
\begin{equation}
\rho_0(y)=e\int d\epsilon\nu(\epsilon)\mbox{sign}(\epsilon)
F(\{\epsilon-\bar{E}_0(y)\}\mbox{sign}(\epsilon)),
\label{charge density}
\end{equation}
where $F(\epsilon)=1/(\exp (\epsilon/k_{\rm B} T)+1)$ is the Fermi distribution function.
Equation (\ref{charge density}) is valid if $\bar{E}_0(y)$ varies slowly on the scale of the Fermi wavelength.

The density of states $\nu$ in semiconducting CNs is given by,\cite{Ajiki_and_Ando_1993a}
\begin{equation}
\nu(\epsilon)=\frac{4}{\pi}\sum_n \frac{|\epsilon|/\gamma}{\sqrt{\epsilon^2-(\gamma\kappa_n)^2}} \theta(|\epsilon|-|\gamma\kappa_n|),
\end{equation}
with the step function $\theta$, the discretized wave number
\begin{equation}
\kappa_n=\frac{1}{R}\left(n-\frac{1}{3}\right),
\end{equation}
the band parameter $\gamma=\sqrt{3}a\gamma_0/2$, and the transfer integral $\gamma_0$.\cite{Lemay_et_al_2001a}

Due to the conservation of the total electron energy the charge neutrality level is related to the electrostatic potential Eq. (\ref{potential}),
\begin{equation}
\bar{E}_0(y)+e\phi(y)=\Delta W,
\label{Energy conservation}
\end{equation}
where $\Delta W$ is a constant and one of the fitting parameters.
Note that, although $\Delta W$ relates to the work function difference between bias electrodes and CNs, it strongly depends on the surface and randomness near the contact.

We solve Eqs. (\ref{potential}), (\ref{charge density}), and (\ref{Energy conservation}) self--consistently to obtain the potential $\phi(y)$ with a boundary condition $\phi(0)=0$.
In order to ensure the boundary condition, we use antisymmetric sources
\begin{equation}
\Psi(y)=V_g \mbox{sign}(y),
\end{equation}
and $\Delta W \mbox{sign}(y)$ instead of the right hand side of Eq. (\ref{Energy conservation}).
For this purpose, image charges at the opposite side of the interface between CN and the metallic electrode have been added in Eq. (\ref{Total Charge Density}).

Transmission coefficients for a single barrier are calculated in the WKB approximation.
The transmission coefficient $T_n$ to the states of the $n$--th band in the CN is
given by
\begin{equation}
T_n(E)=\exp\left[-\int |\kappa_0|\sqrt{\left(\frac{\kappa_n}{\kappa_0}\right)^2-\left(\frac{E-\bar{E}_0 (y)}{\epsilon (0)}\right)^2} dy\right],
\label{Transmission Probability}
\end{equation}
where the half of the band gap $\epsilon(0)=\gamma/3R$, the integral is taken in the barrier, and the energy $E$ is measured from the Fermi energy.
The calculations show that the contribution of the higher sub--bands is negligible compared with that of the lowest band, because of the effectively huge barriers for these sub--bands.
For small bias voltage $V$, the current is given by the Landauer formula at finite temperature by the use of the calculated potential.
A correction for impurity scattering is ignored, since the transmission probability in the CN is much larger than the tunneling probability at the Schottky barrier.
We can ignore interference within CNs at room temperature, and the use Ohm's law for the series resistance of the two barriers near source and drain electrodes.

We now calculate the transistor characteristics for the CN devices.
In the following numerical calculations, a number of parameters are fixed at the experimental values.
The energy gap is fixed at $2\epsilon(0)=0.7$ eV, which is the average value measured on nanotubes with STM.\cite{Venema_et_al_2000a}
By taking the experimentally deduced value $\gamma_0=2.6$ eV,\cite{Lemay_et_al_2001a} this results in a radius $R=0.53$ nm.
The calculations are done for a nanotube with a length of $A=380R=200$ nm and surrounded in a uniform dielectric with $\kappa=5$.
The bias voltage is set at $V=5$ mV and the temperature at $T=300$ K.
Note that the horizontal position of the $I(V_g)$ curve is a free parameter in our fit.
We make this choice because experimentally the whole $I(V_g)$ curve can be shifted horizontally in a hysteresis way when voltages as large as 4V are applied on the gate.
This hysteresis is also observed in samples with a Si gate and possibly originates from trapped oxide charges within the insulators.

Figures \ref{Model} (a), (b) and (c) show that the calculated $I(V_g)$ are in good agreement with the measurements.
The model reproduces the gap as well as the asymmetry on the p and the n doping regions.
In the fit of our model we have used four parameters: $R_s$, $\Delta W$, $f$ and $D_0$.
$R_s$ and $\Delta W$ are the most important parameters.
The gap depends principally on $R_s$ and the asymmetry on $\Delta W$.
The variation of the parameters $f$ and $D_0$ is used for the fine tuning.
The parameter $f$ has an influence on the overall slope of the curves, and $D_0$ on the slope far from the gap.
The values of the fitting parameters are given in the inset of Fig.\ \ref{Model}.

\begin{figure}[t]
\centerline{\epsfig{figure=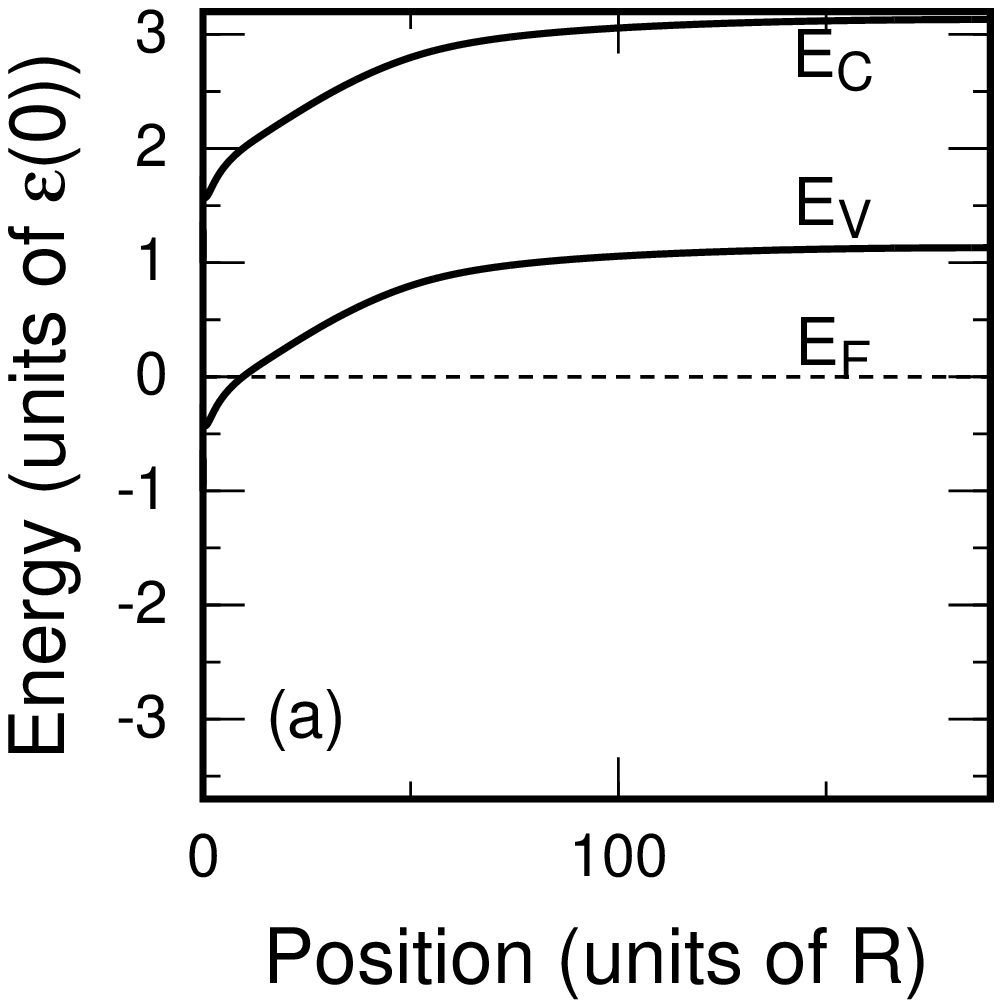,width=4.0cm}\epsfig{figure=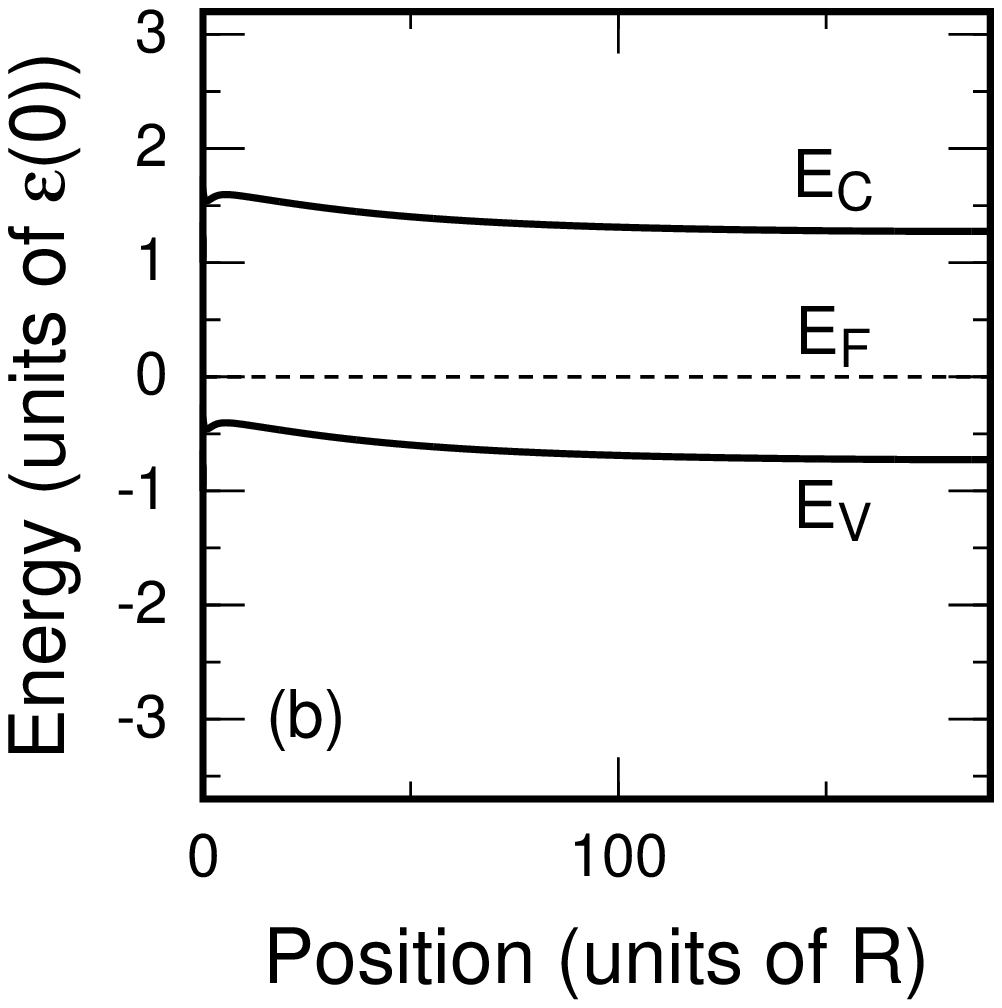,width=4.0cm}}
\centerline{\epsfig{figure=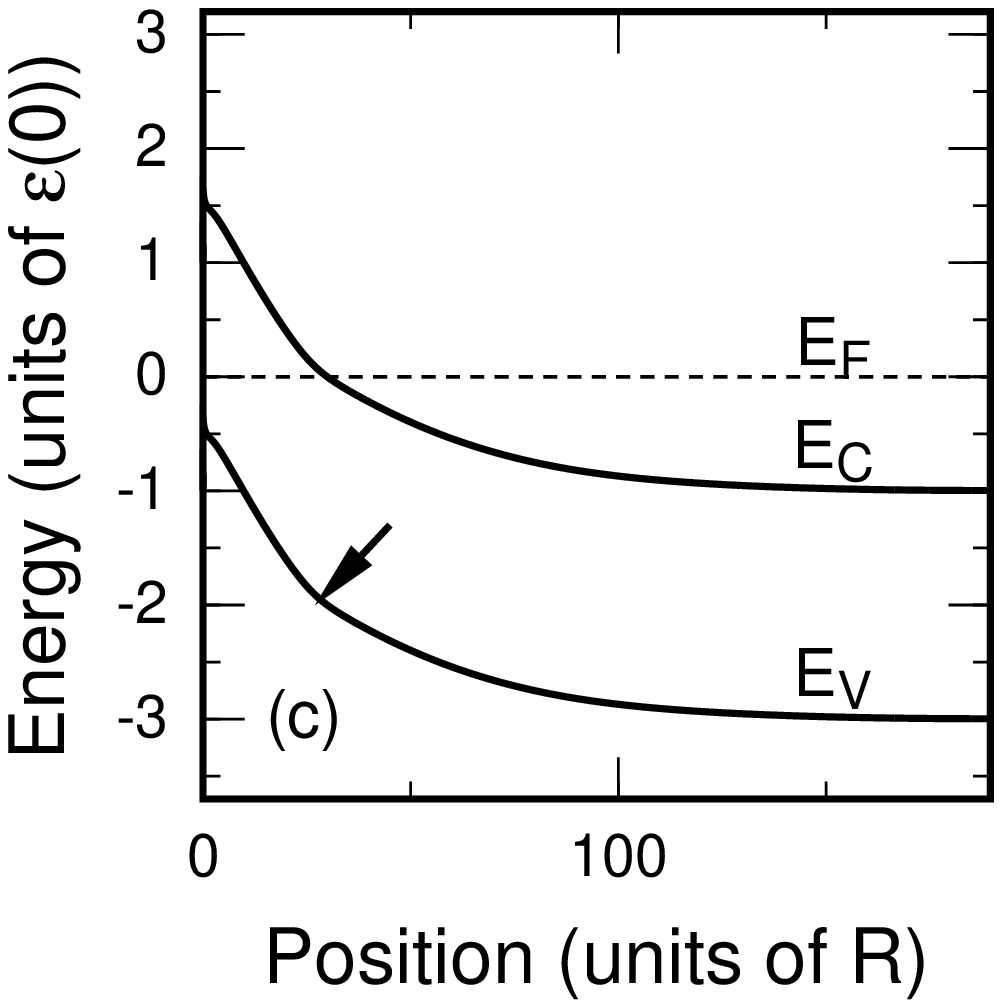,width=4.0cm}\epsfig{figure=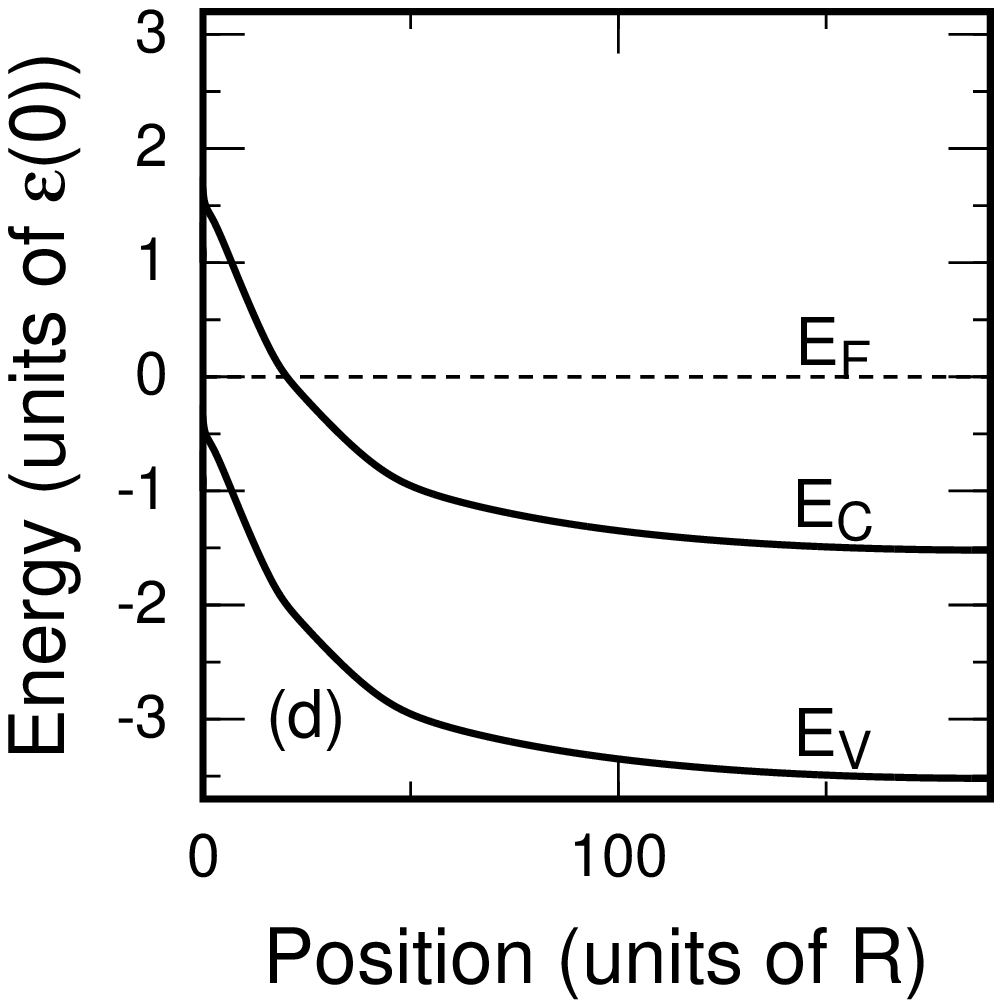,width=4.0cm}}
\caption{Calculated bottom of the conduction band and top of the valence band from the contact to the middle of CN for gate voltages $V_g=-2.3$ (a), $-0.5$ (b), $1.2$ (c), and $2.1$ (d).
The corresponding current is shown in Fig. \ref{Work-function difference} of $\Delta W=0.65\epsilon(0)$.
The arrow in (c) illustrate the effect of a higher subband.
 \label{charge-neutrality level}}
\end{figure}
Figure \ref{charge-neutrality level} shows the band bending as a function of the position from the electrode to the middle of the CN for different gate voltages.
The bands have been calculated with the parameters used for the fit in Fig.\ \ref{Model} (a).
The different band diagrams show that $V_g$ successively shifts the Fermi energy from the valence band to the gap, and then to the conduction band.
It is thus possible to electrostatically change the doping of the nanotube over the full range from p to n doping.

The Schottky barrier height is not expected to be pinned at the middle of the gap, as it is usually the case in 3d, and is that expected for an unpinned junction.
The reason is that, as has been shown theoretically, the effect of the metal--induced gap states\cite{Tersoff_1984a} is limited, because they are on the ring at the contact, instead of a plane for a conventional bulk junction.\cite{Leonard_and_Tersoff_2000a}
Here, the barrier height is equal to $\epsilon(0)-\Delta W$ for p--doped CNs [Fig. \ref{charge-neutrality level} (a)], and $\epsilon(0)+\Delta W$ for n--doped [Fig. \ref{charge-neutrality level} (c) and (d)].
The barrier is lower and thinner  in Fig. \ref{charge-neutrality level} (a) than (c) and (d), because the Fermi level of the electrode is close to the top of valence band.
This gives the asymmetry in the $I(V_g)$ curves between the two regions.

Figure \ref{charge-neutrality level} shows also that the band bending varies very smoothly in space.
The strongest band bending variation occurs at a length scale of the order of the separation between the CN and the gate, in agreement with the calculations where the gate is modelled by a coaxial metal.\cite{Odintsov_2000a,Odintsov_and_Tokura_2000a}
Superposed on this variation, weak kinks can be observed.
These kinks are situated at the integer multiple of $\epsilon(0)$; an example is indicated by the arrow in Fig.\ \ref{charge-neutrality level} (c).
After the next sub--band is occupied, the screening becomes stronger.
The kinks are due to the divergence of the density of state at the sub--band edges.

\begin{figure}[t]
\centerline{\epsfig{figure=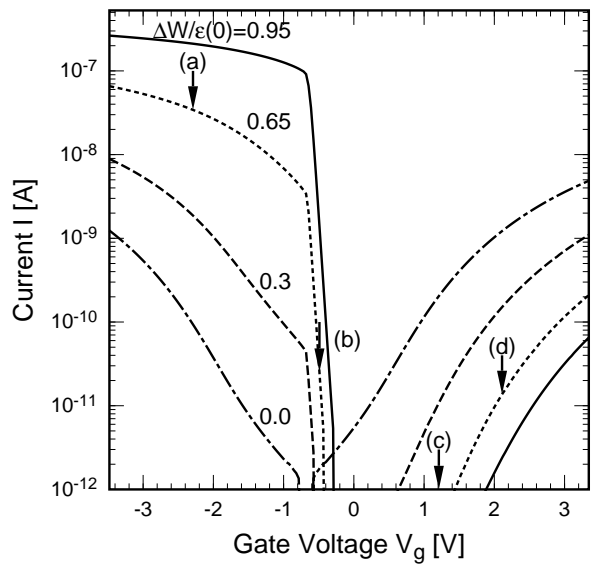,width=8.0cm}}
\caption{$\Delta W$ dependence of the current for $R_s/R=48$ and $D_0=0.1$.
The experiment shown in \ref{Model} (a) can be fitted by the calculation for $\Delta W=0.65\epsilon(0)$.
\label{Work-function difference}}
\end{figure}
Fig.\ \ref{Work-function difference} shows the variation of $I(V_g)$ for different $\Delta W$ values.
The parameters $R_s$, $f$ and $D_0$ are taken equal to what was obtained in Fig.\ \ref{Model} (a).
The arrows (a),(b),(c) and (d) indicate the gate voltages for which the band diagram was represented in  Fig.\ \ref{charge-neutrality level}.
The current is surprisingly larger for situations (b) than for (c) and (d).
Although (b) corresponds to the case that the Fermi level lies in the gap, a non-zero current passes through the tube due to the broadening of the Fermi distribution function at 300 K.
In (c), the Fermi level lies in the conduction band, but the large barrier hardly allows for electrons to tunnel.
The current therefore is lower than $10^{-12}$ A, the lower limit in the measurements.

The calculated current is also shown for other $\Delta W$.
The on--off ratio increases with $\Delta W$, because the Schottky barrier is thin and low for large $\Delta W$.
This indicates that the Fermi energy should be located as close as possible to the valence band edge in order to get larger on--off ratios.
The maximum current and the asymmetry also increase with $\Delta W$.
If $\Delta W$ is negative, the calculated current is the same after the mirror inversion of gate voltage around $V_g=-0.68$ V.

\begin{figure}[t]
\centerline{\epsfig{figure=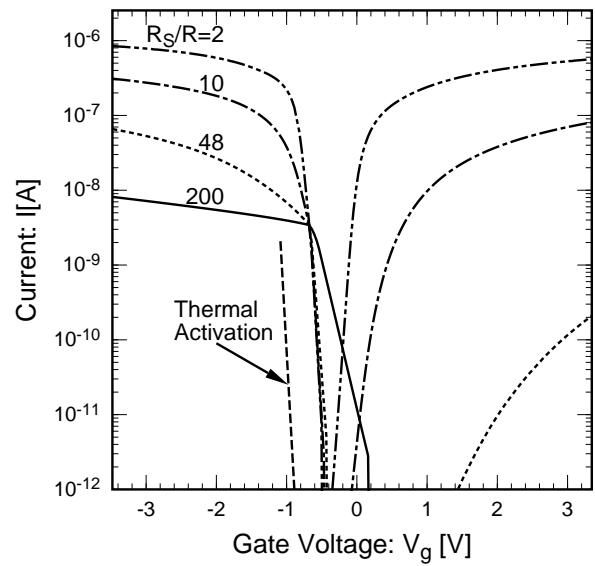,width=8.0cm}}
\caption{Calculated current for several gate distance $R_s$, and $\Delta W=0.65\epsilon(0)$, $D_0=0.1$.
The dashed line on the left shows the thermal--activation slope of $-e/k_{\rm B} T$.
\label{Gate effect}}
\end{figure}
Figure \ref{Gate effect} shows the dependence of the $I(V_g)$ curve on the gate distance $R_s$.
These calculations show the importance of the gate--CN separation for the on-off ratio.
For thinner insulator thickness, a larger on--off ratio is obtained.
Indeed, a thin insulator thickness gives a thin barrier width and thus a larger current.
For the same reason the gap increases with $R_s$.
The current corresponding to the conduction band is out of range in Fig.\ \ref{Gate effect} for $R_s=200 R$.
A very large gate voltage $V_g=13.4$ V is needed to obtain a current of $10^{-12}$ A.
In previous experiments, the separation between the CN and the gate was large. It indeed did not allow to observe a current corresponding in the n-doping region.\cite{Tans_et_al_1998a,Martel_et_al_1998a}

Further calculations (not presented here) show that the slopes around $V_g=-0.5$ V depend on temperature, but poorly on parameters like $R_s$, $\Delta W$ or $D_0$ (for small enough $R_s$, say $R_s \le 100 R$).
The temperature dependence in this $V_g$ region originates from the broadening of the Fermi distribution function at room temperature.
The current can be approximated in a thermally activated form  $I\propto\exp[-eV_g/k_B T]$ with Boltzmann constant $k_B$, as shown in the figure by a dashed line.
For small $R_s$ values, the band bending is mainly located near the interface.
This changes for large $R_s$.
Then, the band bending variation becomes comparable to the nanotube length and the exponential dependence changes.

The separation $R_s$ between the CN and the gate as found in the fit appears to be larger than $48R$.
This value is in relatively poor agreement with the experimental value that is of order $10R$.
This discrepancy probably originates in the description of the bias electrodes.
In the experiment, the electrodes may better screen the electric fields from the gate to the CN than in the model.
The assumption that the electrons tunnel in the end of the CN may also be too strong.
The electrons are in fact likely to tunnel into the CN at a point that is not right at the electrode edge, but close to it.

In conclusion, we have studied the tunnel conductance through the Schottky barrier at the metal--CNs interface with a planar gate model.
Good agreement with the experiment is obtained for several samples.
The asymmetry between the p--doped and the n--doped region has been shown to depend mainly on the difference between the electrode Fermi level and the band gap of carbon nanotubes.
We have shown that the band bending variation in the nanotube length is long.
We have also shown that the choice of source and drain electrodes is important in order to obtain a large on--off ratio.
We have explained the bend of the current versus gate voltage as the transition from a thermal--activation region to a tunneling region.

The authors wish to thank T. Ando, G. E. W. Bauer, Ya. M. Blanter, and P. Hadley for helpful discussions.
The research has been supported by the European Community SATURN project and by the Dutch Foundation for Fundamental Research on Material (FOM).
One of us (T.N.) acknowledges the support of JSPS Postdoctoral Fellowships for Research Abroad.

\def\PRB{Phys.\ Rev.\ B}
\def\PRL{Phys.\ Rev.\ Lett.\null}
\def\APL{Appl.\ Phys.\ Lett.\null}
\def\JPSJ{J.\ Phys.\ Soc.\ Jpn.\null}

\bibliography{nssbe}

\begin{thebibliography}{28}
\expandafter\ifx\csname natexlab\endcsname\relax\def\natexlab#1{#1}\fi
\expandafter\ifx\csname bibnamefont\endcsname\relax
  \def\bibnamefont#1{#1}\fi
\expandafter\ifx\csname bibfnamefont\endcsname\relax
  \def\bibfnamefont#1{#1}\fi
\expandafter\ifx\csname citenamefont\endcsname\relax
  \def\citenamefont#1{#1}\fi
\expandafter\ifx\csname url\endcsname\relax
  \def\url#1{\texttt{#1}}\fi
\expandafter\ifx\csname urlprefix\endcsname\relax\def\urlprefix{URL }\fi
\providecommand{\bibinfo}[2]{#2}
\providecommand{\eprint}[2][]{\url{#2}}

\bibitem[{\citenamefont{Bachtold et~al.}(2001)\citenamefont{Bachtold, Hadley,
  Nakanishi, and Dekker}}]{Bachtold_et_al_2001a}
\bibinfo{author}{\bibfnamefont{A.}~\bibnamefont{Bachtold}},
  \bibinfo{author}{\bibfnamefont{P.}~\bibnamefont{Hadley}},
  \bibinfo{author}{\bibfnamefont{T.}~\bibnamefont{Nakanishi}},
  \bibnamefont{and} \bibinfo{author}{\bibfnamefont{C.}~\bibnamefont{Dekker}},
  \bibinfo{journal}{Science} \textbf{\bibinfo{volume}{294}},
  \bibinfo{pages}{1317} (\bibinfo{year}{2001}).

\bibitem[{\citenamefont{Iijima}(1991)}]{Iijima_1991a}
\bibinfo{author}{\bibfnamefont{S.}~\bibnamefont{Iijima}},
  \bibinfo{journal}{Nature} \textbf{\bibinfo{volume}{354}}, \bibinfo{pages}{56}
  (\bibinfo{year}{1991}).

\bibitem[{\citenamefont{Iijima and Ichihashi}(1993)}]{Iijima_et_al_1993a}
\bibinfo{author}{\bibfnamefont{S.}~\bibnamefont{Iijima}} \bibnamefont{and}
  \bibinfo{author}{\bibfnamefont{T.}~\bibnamefont{Ichihashi}},
  \bibinfo{journal}{Nature} \textbf{\bibinfo{volume}{363}},
  \bibinfo{pages}{603} (\bibinfo{year}{1993}).

\bibitem[{\citenamefont{Bethune et~al.}(1993)\citenamefont{Bethune, Kiang,
  de~Vries, Gorman, Savoy, Vazquez, and Beyers}}]{Bethune_et_al_1993a}
\bibinfo{author}{\bibfnamefont{D.~S.} \bibnamefont{Bethune}},
  \bibinfo{author}{\bibfnamefont{C.~H.} \bibnamefont{Kiang}},
  \bibinfo{author}{\bibfnamefont{M.~S.} \bibnamefont{de~Vries}},
  \bibinfo{author}{\bibfnamefont{G.}~\bibnamefont{Gorman}},
  \bibinfo{author}{\bibfnamefont{R.}~\bibnamefont{Savoy}},
  \bibinfo{author}{\bibfnamefont{J.}~\bibnamefont{Vazquez}}, \bibnamefont{and}
  \bibinfo{author}{\bibfnamefont{R.}~\bibnamefont{Beyers}},
  \bibinfo{journal}{Nature} \textbf{\bibinfo{volume}{363}},
  \bibinfo{pages}{605} (\bibinfo{year}{1993}).

\bibitem[{\citenamefont{Kong et~al.}(1999)\citenamefont{Kong, Han, and
  Ihm}}]{Kong_et_al_1999a}
\bibinfo{author}{\bibfnamefont{K.}~\bibnamefont{Kong}},
  \bibinfo{author}{\bibfnamefont{S.}~\bibnamefont{Han}}, \bibnamefont{and}
  \bibinfo{author}{\bibfnamefont{J.}~\bibnamefont{Ihm}},
  \bibinfo{journal}{\PRB} \textbf{\bibinfo{volume}{60}}, \bibinfo{pages}{6074}
  (\bibinfo{year}{1999}).

\bibitem[{\citenamefont{Choi et~al.}(1999)\citenamefont{Choi, Ihm, Yoon, and
  Louie}}]{Choi_et_al_1999a}
\bibinfo{author}{\bibfnamefont{H.~J.} \bibnamefont{Choi}},
  \bibinfo{author}{\bibfnamefont{J.}~\bibnamefont{Ihm}},
  \bibinfo{author}{\bibfnamefont{Y.~G.} \bibnamefont{Yoon}}, \bibnamefont{and}
  \bibinfo{author}{\bibfnamefont{S.~G.} \bibnamefont{Louie}},
  \bibinfo{journal}{\PRB} \textbf{\bibinfo{volume}{60}},
  \bibinfo{pages}{R14009} (\bibinfo{year}{1999}).

\bibitem[{\citenamefont{Rochefort et~al.}(2001)\citenamefont{Rochefort, Ventra,
  and Avouris}}]{Rochefort_et_al_2001a}
\bibinfo{author}{\bibfnamefont{A.}~\bibnamefont{Rochefort}},
  \bibinfo{author}{\bibfnamefont{M.~D.} \bibnamefont{Ventra}},
  \bibnamefont{and} \bibinfo{author}{\bibfnamefont{P.}~\bibnamefont{Avouris}},
  \bibinfo{journal}{\APL} \textbf{\bibinfo{volume}{78}}, \bibinfo{pages}{2521}
  (\bibinfo{year}{2001}).

\bibitem[{\citenamefont{Anantram et~al.}(2000)\citenamefont{Anantram, Datta,
  and Xue}}]{Anantram_et_al_2000a}
\bibinfo{author}{\bibfnamefont{M.~P.} \bibnamefont{Anantram}},
  \bibinfo{author}{\bibfnamefont{S.}~\bibnamefont{Datta}}, \bibnamefont{and}
  \bibinfo{author}{\bibfnamefont{Y.}~\bibnamefont{Xue}},
  \bibinfo{journal}{\PRB} \textbf{\bibinfo{volume}{61}}, \bibinfo{pages}{14219}
  (\bibinfo{year}{2000}).

\bibitem[{\citenamefont{Nakanishi and Ando}(2000)}]{Nakanishi_and_Ando_2000a}
\bibinfo{author}{\bibfnamefont{T.}~\bibnamefont{Nakanishi}} \bibnamefont{and}
  \bibinfo{author}{\bibfnamefont{T.}~\bibnamefont{Ando}},
  \bibinfo{journal}{\JPSJ} \textbf{\bibinfo{volume}{69}}, \bibinfo{pages}{2175}
  (\bibinfo{year}{2000}).

\bibitem[{\citenamefont{Odintsov}(2000)}]{Odintsov_2000a}
\bibinfo{author}{\bibfnamefont{A.~A.} \bibnamefont{Odintsov}},
  \bibinfo{journal}{\PRL} \textbf{\bibinfo{volume}{85}}, \bibinfo{pages}{150}
  (\bibinfo{year}{2000}).

\bibitem[{\citenamefont{Odintsov and Tokura}(2000)}]{Odintsov_and_Tokura_2000a}
\bibinfo{author}{\bibfnamefont{A.~A.} \bibnamefont{Odintsov}} \bibnamefont{and}
  \bibinfo{author}{\bibnamefont{Tokura}}, \bibinfo{journal}{J.\ Low\ Temp.\
  Phys.} \textbf{\bibinfo{volume}{118}}, \bibinfo{pages}{509}
  (\bibinfo{year}{2000}).

\bibitem[{\citenamefont{L\'eonard and
  Tersoff}(1999)}]{Leonard_and_Tersoff_1999a}
\bibinfo{author}{\bibfnamefont{F.}~\bibnamefont{L\'eonard}} \bibnamefont{and}
  \bibinfo{author}{\bibfnamefont{J.}~\bibnamefont{Tersoff}},
  \bibinfo{journal}{\PRL} \textbf{\bibinfo{volume}{83}}, \bibinfo{pages}{5174}
  (\bibinfo{year}{1999}).

\bibitem[{\citenamefont{L\'eonard and
  Tersoff}(2000)}]{Leonard_and_Tersoff_2000a}
\bibinfo{author}{\bibfnamefont{F.}~\bibnamefont{L\'eonard}} \bibnamefont{and}
  \bibinfo{author}{\bibfnamefont{J.}~\bibnamefont{Tersoff}},
  \bibinfo{journal}{\PRL} \textbf{\bibinfo{volume}{84}}, \bibinfo{pages}{4693}
  (\bibinfo{year}{2000}).

\bibitem[{\citenamefont{Tersoff}(1999)}]{Tersoff_1999a}
\bibinfo{author}{\bibfnamefont{J.}~\bibnamefont{Tersoff}},
  \bibinfo{journal}{\APL} \textbf{\bibinfo{volume}{74}}, \bibinfo{pages}{2122}
  (\bibinfo{year}{1999}).

\bibitem[{\citenamefont{Xue and Datta}(1999)}]{Xue_and_Datta_1999a}
\bibinfo{author}{\bibfnamefont{Y.}~\bibnamefont{Xue}} \bibnamefont{and}
  \bibinfo{author}{\bibfnamefont{S.}~\bibnamefont{Datta}},
  \bibinfo{journal}{\PRL} \textbf{\bibinfo{volume}{83}}, \bibinfo{pages}{4844}
  (\bibinfo{year}{1999}).

\bibitem[{\citenamefont{Tans et~al.}(1998)\citenamefont{Tans, Verschueren, and
  Dekker}}]{Tans_et_al_1998a}
\bibinfo{author}{\bibfnamefont{S.~J.} \bibnamefont{Tans}},
  \bibinfo{author}{\bibfnamefont{A.~R.~M.} \bibnamefont{Verschueren}},
  \bibnamefont{and} \bibinfo{author}{\bibfnamefont{C.}~\bibnamefont{Dekker}},
  \bibinfo{journal}{Nature} \textbf{\bibinfo{volume}{393}}, \bibinfo{pages}{49}
  (\bibinfo{year}{1998}).

\bibitem[{\citenamefont{Martel et~al.}(1998)\citenamefont{Martel, Schmidt,
  Shea, Hertel, and Avouris}}]{Martel_et_al_1998a}
\bibinfo{author}{\bibfnamefont{R.}~\bibnamefont{Martel}},
  \bibinfo{author}{\bibfnamefont{T.}~\bibnamefont{Schmidt}},
  \bibinfo{author}{\bibfnamefont{H.~R.} \bibnamefont{Shea}},
  \bibinfo{author}{\bibfnamefont{T.}~\bibnamefont{Hertel}}, \bibnamefont{and}
  \bibinfo{author}{\bibfnamefont{P.}~\bibnamefont{Avouris}},
  \bibinfo{journal}{\APL} \textbf{\bibinfo{volume}{73}}, \bibinfo{pages}{2447}
  (\bibinfo{year}{1998}).

\bibitem[{\citenamefont{Zhou et~al.}(2000)\citenamefont{Zhou, Kong, and
  Dai}}]{Zhou_et_al_2000a}
\bibinfo{author}{\bibfnamefont{C.}~\bibnamefont{Zhou}},
  \bibinfo{author}{\bibfnamefont{J.}~\bibnamefont{Kong}}, \bibnamefont{and}
  \bibinfo{author}{\bibfnamefont{H.}~\bibnamefont{Dai}},
  \bibinfo{journal}{\APL} \textbf{\bibinfo{volume}{76}}, \bibinfo{pages}{1597}
  (\bibinfo{year}{2000}).

\bibitem[{\citenamefont{Freitag et~al.}(2001)\citenamefont{Freitag,
  Radosavljevic, Zhou, Johnson, and Smith}}]{Freitag_et_al_2001a}
\bibinfo{author}{\bibfnamefont{M.}~\bibnamefont{Freitag}},
  \bibinfo{author}{\bibfnamefont{M.}~\bibnamefont{Radosavljevic}},
  \bibinfo{author}{\bibfnamefont{Y.}~\bibnamefont{Zhou}},
  \bibinfo{author}{\bibfnamefont{A.~T.} \bibnamefont{Johnson}},
  \bibnamefont{and} \bibinfo{author}{\bibfnamefont{W.~F.} \bibnamefont{Smith}},
  \bibinfo{journal}{\APL} \textbf{\bibinfo{volume}{79}}, \bibinfo{pages}{3326}
  (\bibinfo{year}{2001}).

\bibitem[{\citenamefont{Park and McEuen}(2001)}]{Park_and_Mceuen_2001a}
\bibinfo{author}{\bibfnamefont{J.}~\bibnamefont{Park}} \bibnamefont{and}
  \bibinfo{author}{\bibfnamefont{P.~L.} \bibnamefont{McEuen}},
  \bibinfo{journal}{\APL} \textbf{\bibinfo{volume}{79}}, \bibinfo{pages}{1363}
  (\bibinfo{year}{2001}).

\bibitem[{\citenamefont{Javey et~al.}(2002)\citenamefont{Javey, Shim, and
  Dai}}]{Javey_et_al_2002a}
\bibinfo{author}{\bibfnamefont{A.}~\bibnamefont{Javey}},
  \bibinfo{author}{\bibfnamefont{M.}~\bibnamefont{Shim}}, \bibnamefont{and}
  \bibinfo{author}{\bibfnamefont{H.}~\bibnamefont{Dai}},
  \bibinfo{journal}{\APL} \textbf{\bibinfo{volume}{80}}, \bibinfo{pages}{1064}
  (\bibinfo{year}{2002}).

\bibitem[{\citenamefont{Derycke et~al.}(2001)\citenamefont{Derycke, Martel,
  Appenzeller, and Avouris}}]{Derycke_et_al_2001a}
\bibinfo{author}{\bibfnamefont{V.}~\bibnamefont{Derycke}},
  \bibinfo{author}{\bibfnamefont{R.}~\bibnamefont{Martel}},
  \bibinfo{author}{\bibfnamefont{J.}~\bibnamefont{Appenzeller}},
  \bibnamefont{and} \bibinfo{author}{\bibfnamefont{P.}~\bibnamefont{Avouris}},
  \bibinfo{journal}{Nano Lett} \textbf{\bibinfo{volume}{1}},
  \bibinfo{pages}{453} (\bibinfo{year}{2001}).

\bibitem[{\citenamefont{Martel et~al.}(2001)\citenamefont{Martel, Derycke,
  Lavoie, Appenzeller, Chan, Tersoff, and Avouris}}]{Martel_et_al_2001a}
\bibinfo{author}{\bibfnamefont{R.}~\bibnamefont{Martel}},
  \bibinfo{author}{\bibfnamefont{V.}~\bibnamefont{Derycke}},
  \bibinfo{author}{\bibfnamefont{C.}~\bibnamefont{Lavoie}},
  \bibinfo{author}{\bibfnamefont{J.}~\bibnamefont{Appenzeller}},
  \bibinfo{author}{\bibfnamefont{K.~K.} \bibnamefont{Chan}},
  \bibinfo{author}{\bibfnamefont{J.}~\bibnamefont{Tersoff}}, \bibnamefont{and}
  \bibinfo{author}{\bibfnamefont{P.}~\bibnamefont{Avouris}},
  \bibinfo{journal}{\PRL} \textbf{\bibinfo{volume}{87}},
  \bibinfo{pages}{256805} (\bibinfo{year}{2001}).

\bibitem[{\citenamefont{Bockrath et~al.}(1999)\citenamefont{Bockrath, Cobden,
  Lu, Rinzler, Smalley, Balents, and McEuen}}]{Bockrath_et_al_1999a}
\bibinfo{author}{\bibfnamefont{M.}~\bibnamefont{Bockrath}},
  \bibinfo{author}{\bibfnamefont{D.~H.} \bibnamefont{Cobden}},
  \bibinfo{author}{\bibfnamefont{J.}~\bibnamefont{Lu}},
  \bibinfo{author}{\bibfnamefont{A.~G.} \bibnamefont{Rinzler}},
  \bibinfo{author}{\bibfnamefont{R.~E.} \bibnamefont{Smalley}},
  \bibinfo{author}{\bibfnamefont{L.}~\bibnamefont{Balents}}, \bibnamefont{and}
  \bibinfo{author}{\bibfnamefont{P.~L.} \bibnamefont{McEuen}},
  \bibinfo{journal}{Nature} \textbf{\bibinfo{volume}{397}},
  \bibinfo{pages}{598} (\bibinfo{year}{1999}).

\bibitem[{\citenamefont{Ajiki and Ando}(1993)}]{Ajiki_and_Ando_1993a}
\bibinfo{author}{\bibfnamefont{H.}~\bibnamefont{Ajiki}} \bibnamefont{and}
  \bibinfo{author}{\bibfnamefont{T.}~\bibnamefont{Ando}},
  \bibinfo{journal}{\JPSJ} \textbf{\bibinfo{volume}{62}}, \bibinfo{pages}{1255}
  (\bibinfo{year}{1993}).

\bibitem[{\citenamefont{Lemay et~al.}(2001)\citenamefont{Lemay, Janssen,
  van~den Hout, Mooij, Broikowski, Willis, Smalley, Kouwenhoven, and
  Dekker}}]{Lemay_et_al_2001a}
\bibinfo{author}{\bibfnamefont{S.~G.} \bibnamefont{Lemay}},
  \bibinfo{author}{\bibfnamefont{J.~W.} \bibnamefont{Janssen}},
  \bibinfo{author}{\bibfnamefont{M.}~\bibnamefont{van~den Hout}},
  \bibinfo{author}{\bibfnamefont{M.}~\bibnamefont{Mooij}},
  \bibinfo{author}{\bibfnamefont{M.~J.} \bibnamefont{Broikowski}},
  \bibinfo{author}{\bibfnamefont{P.~A.} \bibnamefont{Willis}},
  \bibinfo{author}{\bibfnamefont{R.~E.} \bibnamefont{Smalley}},
  \bibinfo{author}{\bibfnamefont{L.~P.} \bibnamefont{Kouwenhoven}},
  \bibnamefont{and} \bibinfo{author}{\bibfnamefont{C.}~\bibnamefont{Dekker}},
  \bibinfo{journal}{Nature} \textbf{\bibinfo{volume}{412}},
  \bibinfo{pages}{617} (\bibinfo{year}{2001}).

\bibitem[{\citenamefont{Venema et~al.}(2000)\citenamefont{Venema, Janssen, and
  Buitelaar}}]{Venema_et_al_2000a}
\bibinfo{author}{\bibfnamefont{L.~C.} \bibnamefont{Venema}},
  \bibinfo{author}{\bibfnamefont{J.~W.} \bibnamefont{Janssen}},
  \bibnamefont{and} \bibinfo{author}{\bibfnamefont{M.~R.}
  \bibnamefont{Buitelaar}}, \bibinfo{journal}{\PRB}
  \textbf{\bibinfo{volume}{62}}, \bibinfo{pages}{5238} (\bibinfo{year}{2000}).

\bibitem[{\citenamefont{Tersoff}(1984)}]{Tersoff_1984a}
\bibinfo{author}{\bibfnamefont{J.}~\bibnamefont{Tersoff}},
  \bibinfo{journal}{\PRL} \textbf{\bibinfo{volume}{52}}, \bibinfo{pages}{465}
  (\bibinfo{year}{1984}).

\end{thebibliography}

\end{document}